# HAD THE PLANET MARS NOT EXISTED: KEPLER'S EQUANT MODEL AND ITS PHYSICAL CONSEQUENCES

C Bracco[1,2] and J-P Provost[3]

[1] UMR Fizeau, Université de Nice-Sophia Antipolis, CNRS, Observatoire de la Côte d'Azur, Campus Valrose, 06108 Nice Cedex, France

[2] Syrte, CNRS, Observatoire de Paris, 61 avenue de l'Observatoire, 75014 PARIS

[3] INLN, Université de Nice-Sophia Antipolis, 1361 route des lucioles, Sophia-Antipolis, 06560 Valbonne, France

**Abstract:** We examine the equant model for the motion of planets, which has been the starting point of Kepler's investigations before he modified it because of Mars observations. We show that, up to first order in eccentricity, this model implies for each orbit a velocity which satisfies Kepler's second law and Hamilton's hodograph, and a centripetal acceleration with an $r^{-2}$ dependence on the distance to the sun. If this dependence is assumed to be universal, Kepler's third law follows immediately. This elementary exercise in kinematics for undergraduates emphasizes the proximity of the equant model coming from Ancient Greece with our present knowledge. It adds to its historical interest a didactical relevance concerning, in particular, the discussion of the Aristotelian or Newtonian conception of motion.

## 1. Introduction.

The introduction of the equant hypothesis in the description of the motion of celestial bodies is considered by historians as a major achievement of mathematical science in Antiquity (see e.g. [1]-[3]). Basically, it consists to consider that the motion of a point $P$ on a circle of center $C$ is a natural one if inside the circle there is a fixed point $Q$, called the equant point, such that the direction $QP$ rotates uniformly. If $Q$ is the center $C$, one recovers the uniform circular motions which were the only ones previously admitted.[1] The extensive use of epicycles in the geocentric system of Ptolemy, which has led him to establish astronomical tables still in use at the beginning of the 17th century, has unfortunately contributed to hide the importance of the equant hypothesis. Even Copernicus rejected it in his heliocentric system.[2] On the contrary, Kepler, who was convinced by heliocentrism, but who also considered the Sun as the "motor" of the planetary motion (see e.g. [4]), took the equant hypothesis as more physically acceptable than that of epicycles, already in his *Mysterium Cosmographicum* [5]. He quickly applied it to the motion of planets, including the Earth.

---

[1] For example, in order to account for the unequal durations of seasons, Hipparchus (2nd century B.C.), who took for granted that the sun $S$ turns uniformly on a circle, was led to assume that the Earth has an eccentric position $E$ so that the apparent direction $ES$ does not rotate uniformly. This model is known as the eccentric model.

[2] The equant was considered by Copernicus as a "monstrous hypothesis", since it did not obey the dogma of a uniform circular motion. Copernicus was himself convinced that one of the major insights of his heliocentric model was that it enabled to get rid of the equant. The price to pay by him was the introduction of numerous epicycles to account for the non uniform motion of planets.

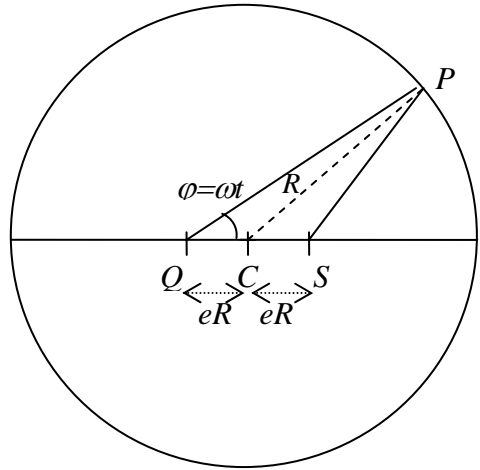

**Fig.1. Definition of the equant model.**

For the sake of definiteness, let us present briefly Kepler's model (figure 1), which can be found in *Astronomia Nova* [6]. The trajectory of the planet $P$ (for example the Earth) is a circle of center $C$ and radius $R$, and the Sun $S$ lies at the distance $eR$ from $C$; $e$ is a small parameter characterizing the eccentricity. The motion of the planet is defined by the uniform rotation of the line $QP$, where the equant point $Q$ is symmetric of $S$ with respect to $C$. In short, Kepler's equant model, is defined by figure 1 and the mathematical relations:

$$QC = CS = eR \quad ; \quad \varphi = \omega t \quad ; \quad \omega = \text{cste}. \tag{1}$$

According to Kepler himself, the equant hypothesis has been "a cornerstone" of his work. Indeed, it is from the equant parameters $e, R$ of the Earth, that he derived accurate distances of Mars to the Sun, and was led to abandon his equant model since these distances did not fit it at the precision of Tycho Brahe's observations.[3] Had Mars be absent from the solar system, or had its eccentricity $e$ be small enough (or the observations be less precise), the equant model would have been for some time the "standard model" for planetary motions.

In the present paper, we discuss Kepler's equant model for planets, not from the practical point of view of its comparison with observations, but from a didactical one, looking at the physical properties or laws which it may imply. In particular, we show that for small eccentricities[4], this model is in conformity with our present knowledge of Keplerian motions:[5] the points $Q$ and $S$ play the role of the focuses of an ellipse, and the motion of $P$ satisfies the law of equal areas; the hodograph of the motion is a circle, as Hamilton proved it in the general case [11]; the acceleration of $P$ is a centripetal one and satisfies Newton's $r^{-2}$ law on the orbit. We also show that if, starting with the equant model, one looks for a general relation (valid for any planet) concerning the dependence with respect to the distance $r$ to the Sun, either of the velocity of the planets (Aristotelian conception of motion), or of their acceleration (Newtonian conception), only the latter leads to the correct law for the orbital periods (Kepler's third law). In our opinion, these results, which are easily accessible to undergraduates, emphasize that Kepler's equant model is interesting, not only from an historical point of view, but also from a didactical one in the teaching of Mechanics. They could also be adapted by secondary school teachers to illustrate the importance of the role played by acceleration in Mechanics.

---

[3] In [6] Kepler determined the orbit of the Earth by fitting a circle to three positions at different Martian years and verified that its motion agrees with the equant hypothesis. Before introducing elliptical trajectories, he also tried to find a non symmetric position of $Q$ for Mars (the "vicaria hypothesis" examined in [7]).

[4] The property of the equant model to be the first approximation in eccentricity of a general Keplerian motion has already been pointed out in the literature (see e.g. [8]), but up to our knowledge, not with pedagogical purposes.

[5] For a simple derivation of Kepler's laws starting from Newton's original analysis see [9]. For a standard course on Mechanics, see e.g. [10].

In the whole paper, we consider that the above defined eccentricity is a small parameter $(e \ll 1)$, and all the calculations are limited to the first order in $e$. After introducing a few useful geometrical and kinematical relations (section 2), we examine successively Kepler's first law (section 3), Kepler's second law and the possibility of a "velocity law" for the planets (section 4), Hamilton's hodograph (section 5), and finally, Kepler's third law in connection with Newton's universal law of gravitation (section 6). In conclusion (section 7), we briefly discuss the interest of the introduction of the equant model in the teaching of Mechanics.

## 2. A few useful relations.

For a mathematical setting of this model, we need to compare the angles $\varphi$, $\theta_0$, $\theta$ of the three directions $QP$, $CP$, $SP$ with the axis $QS$ ($x$-axis), as well as the distances $QP$, $R = CP$ and $r = SP$ (figure 2). The variables $(r, \theta)$ are the polar coordinates of the planet $P$ with respect to the Sun $S$. The origin of the angles corresponds to the perihelion $P_0$.

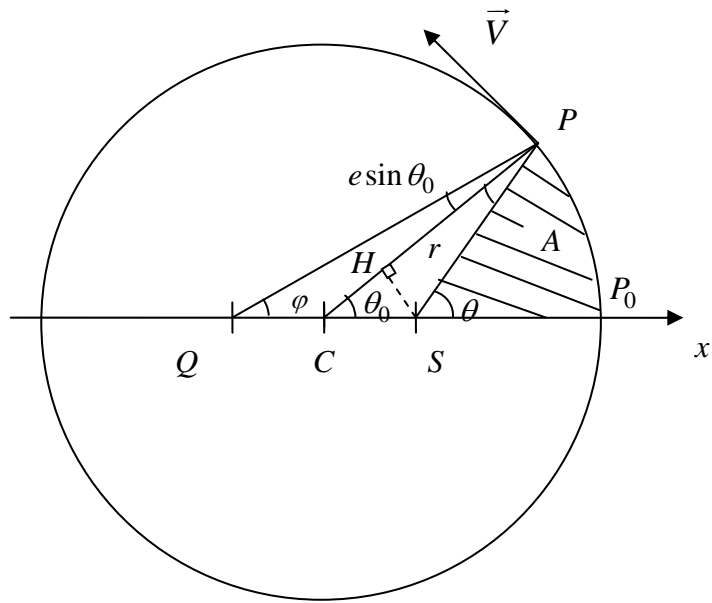

**Fig. 2. Definition of the polar coordinate $r, \theta$, the perihelion $P_0$ and the swept area $A$.** The angles at $P$ are equal to $e \sin \theta_0$.

Projecting the point $S$ in $H$ on the line $CP$ and noting that $CH = eR \cos \theta_0$ and $SH = eR \sin \theta_0$, one gets at first order in $e$:

$$r = R(1 - e\cos\theta_0) \tag{2}$$

$$\theta = \theta_0 + e\sin\theta_0. \tag{3}$$

In equation (3) $e\sin\theta_0$ is the angle at $P$ of the triangle $CPS$. In the same way, it is easy to obtain:

$$QP = R(1 + e\cos\theta_0) = \frac{R^2}{r} \tag{4}$$

$$\varphi = \theta_0 - e\sin\theta_0. \tag{5}$$

As expected, the three angles and three distances differ from each other by first order terms in $e$; this allows replacing indifferently $\theta_0$ by $\theta$ or $\varphi$ in terms proportional to $e$.

From these relations, the speed $V$ of the planet, which is rigorously equal to $R\dot\theta_0$ (the dot meaning time derivative as usual), can also be written:[6]

$$V = r\dot\theta = \omega \times QP. \tag{6}$$

Let us remark that these expressions of $V$, which constitute the basis of the equant kinematics, are also intuitive. Indeed, since the segments $SP$ or $QP$ are almost orthogonal to the circle at $P$, the distance travelled by $P$ in a small time interval is also, up to $e^2$ terms, the distance deduced from rotations around $S$ or $Q$ at the respective angular velocities $\dot\theta$ or $\omega$.

### 3. Kepler's first law.

This law says that the trajectory of a planet is an ellipse with the Sun $S$ at a focus. Then, as well known, the polar coordinates and $r$ and $\theta$ are related by the equation

$$r = \frac{p}{1 + e\cos\theta}, \qquad p = a(1 - e^2), \tag{7}$$

where $e$, $p$ and $a$ are respectively the eccentricity, the parameter and the major semi-axis of the ellipse. It is obvious that this equation (with $p = R$) is in first order in $e$ identical to equation (2) for the distance $SP$ of the planet to the Sun in the equant model.

It may seem at first sight paradoxical, that the circular orbit of the equant model obeys the equation which characterizes an ellipse, because one expects this to be true only for $e = 0$. But this property extends to first order in $e$ since the ratio of the two axis of the ellipse is $\sqrt{1 - e^2}$. The difference between $e = 0$ and $e$ small is that the focuses keep separate at first order in $e$. In the equant model, the point $Q$ plays the role of the second focus, as Kepler noticed in [6] after discovering the exact trajectory.[7] In the next sections, this first order equivalence in $e$ of the equant model with Keplerian ellipses will reveal to be true also at the kinematical level.

### 4. Kepler's second law and the associated "velocity law".

Kepler's second law stipulates that equal areas are swept by the radius vector $\overrightarrow{SP}$ in equal time intervals. It gives an operational rule for the speed at different points of the orbit. $P_0$

---

[6] For example, using equations (2) and (3): $r\dot\theta = R(1 - e\cos\theta_0)\dot\theta_0(1 + e\cos\theta_0) = R\dot\theta_0$ up to $e^2$ terms.

[7] Without calling for equation (7), let us remark that the sum $PQ + PS$ of the distances of the planet to the focuses is a constant $2R$, which is the geometrical definition of an ellipse. In geometrical optics, the symmetric points $S$ and $Q$ are conjugate points if the circle of figure 2 is a mirror (Fermat's principle).

being the perihelion, let us calculate the area $A$ of the angular sector $P_0SP$ (figure 2) from the difference between the area of the angular sector $P_0CP$ and that of the triangle $SCP$:

$$A = \text{area}(P_0CP) - \text{area } A(SCP). \tag{8}$$

The first term reading $R^2\theta_0/2$ and the second one $SH \times CP/2$, one gets

$$A = \frac{R^2}{2}(\theta_0 - e\sin\theta_0) = \frac{R^2\varphi}{2}, \tag{9}$$

the last equality resulting from equation (5). Since $\varphi = \omega t$, the area which is swept by $\overrightarrow{SP}$ in the equant model is proportional to time. We emphasize that this would not have been the case in the (heliocentric) eccentric model where it is the angle $\theta_0$ which is proportional to time. In standard lectures in Mechanics, one usually introduces the constant $K$ which is defined as twice the area swept by unit time.[8] In the equant model, it is equal to:

$$K = 2\dot{A} = \omega R^2. \tag{10}$$

Since $R^2 = r \times QP$ and $V = \omega \times QP$ (equations (4) and (6)), the introduction of this constant allows writing the speed of the planet on the trajectory in the form:

$$V = \frac{K}{r}. \tag{11}$$

The proportionality of $V$ to $r^{-1}$, for the perihelion and aphelion in the equant model, was well known of Ptolemy. It is reasonable to think that Kepler has noted that this proportionality is valid for other points of the trajectory. This may be one reason why he considered that the Sun was responsible for the velocity of the planets and that he took already in [5] this proportionality to be a general law of planet motion (the "velocity law").[9] Kepler was also aware, as he noticed in his later comments on [5], that this law implied the (incorrect) proportionality of the period $T$ to $R^2$. We discuss this "velocity law" further in section 7.

## 5. Hamilton's hodograph.

Let us recall that the hodograph is the curve which is described by the extremity of the vector velocity $\vec{V}$, when its origin is kept fixed. This notion was first introduced by Hamilton [11], who discovered that in the case of Keplerian motions (central forces proportional to $r^{-2}$), the hodograph is rigorously a circle.

In order to show (simply and geometrically) that the hodograph in the equant model is also a circle, it is useful to introduce the vector $\vec{V}^* = \vec{V}_{-\pi/2}$ which is deduced from the velocity at $P$ by a $-\pi/2$ rotation. This vector is parallel to $\overrightarrow{CP}$ and its modulus is $\omega \times QP$ according to (6). Therefore, let $W$ be the point on the orbit (figure 3a) such that $\overrightarrow{QW}$ is parallel to $\overrightarrow{CP}$.[10] Since both segments $QP$ and $QW$ issued from $Q$ are close and almost perpendicular to the circle, one has $QP = QW$ up to $e^2$ terms, and $\vec{V}^*$ reads:

$$\vec{V}^* = \omega \times \overrightarrow{QW} = \omega\left(\overrightarrow{QC} + \overrightarrow{CW}\right). \tag{12}$$

---

[8] In these courses, the constant $K$ is also shown to be equal to $r^2\dot{\theta}$ and $\|\vec{r} \wedge \vec{V}\|$. It is easy to verify it in the equant model, using respectively equation (6) and $\|\vec{r} \wedge \vec{V}\| = rV\cos(\theta - \theta_0) \simeq rV$.

[9] For a discussion on the difference between the area law and the "velocity law", see [12].

[10] $W$ is also such that $\overrightarrow{CW}$ is parallel to $\overrightarrow{SP}$ since the angle of $CW$ with the $x$-axis is $\theta_0 + e\sin\theta_0 = \theta$.

If $Q$ is taken as the origin of $\vec{V}^*/\omega$, the extremity of this vector follows remarkably the same trajectory as the planet (figure 3a). As a consequence, since the velocity $\vec{V}$ is, up to the factor $\omega$, deduced from $\vec{V}^*$ by a $\pi/2$ rotation, the hodograph with $Q$ as origin is also an eccentric circle.

In order to have an explicit expression for $\vec{V}$ itself,[11] we must introduce the unit radial vector $\vec{u}_r$, parallel to $\overrightarrow{SP}$ and $\overrightarrow{CW}$, and the orthoradial vector $\vec{u}_\theta$ deduced from $\vec{u}_r$ by a $\pi/2$ rotation. Then, the rotation by $\pi/2$ of the above vector $\vec{V}^*$ leads to (see note 11):

$$\vec{V} = \omega R\left(\vec{e} + \vec{u}_\theta\right). \tag{13}$$

In (13) the vector $\vec{e}$ (sometimes called the eccentricity vector [13]) is of modulus $e$ and is parallel to the $y$-axis (figure 3b).

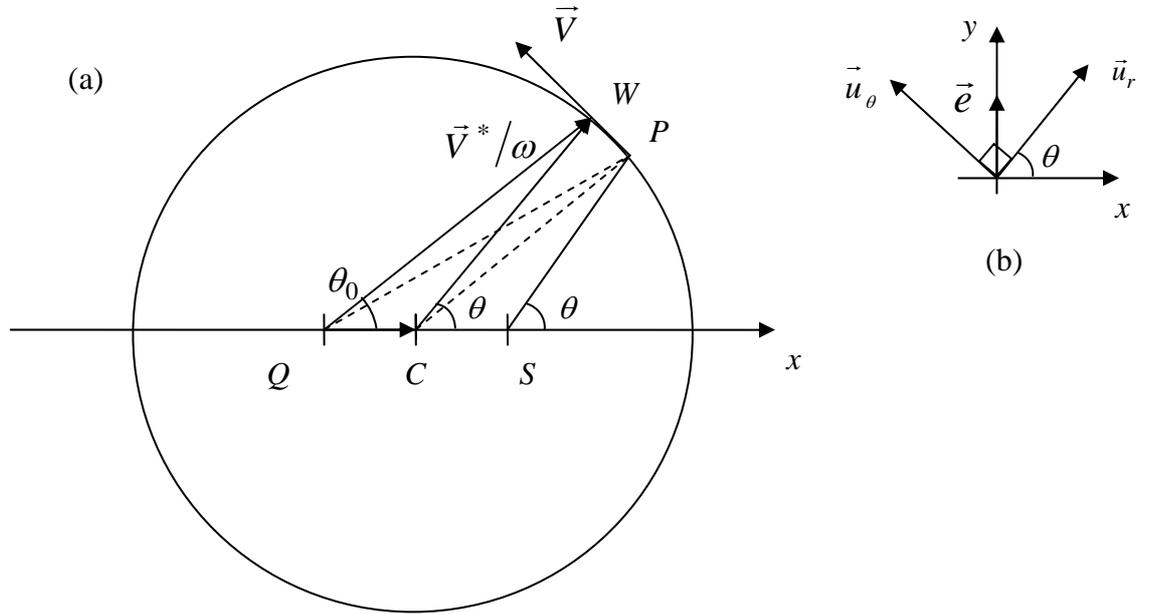

**Fig. 3.** (a) The velocity at $P$ (orthogonal to $CP$) is deduced from $\omega \times \overrightarrow{QW}$ by a $\pi/2$ rotation. $QW$ is parallel to $CP$, and $CW$ to $SP$. (b) Definition of the vectors $\vec{u}_r$, $\vec{u}_\theta$ and $\vec{e}$.

### 6. Newton's law for the acceleration and Kepler's third law.

According to equation (13), the velocity is the sum of a constant vector and of the vector $\omega R \vec{u}_\theta$ which turns at the angular velocity $\dot{\theta}$. Therefore, its derivative (the acceleration) is in the direction of $-\vec{u}_r$ (deduced from $\vec{u}_\theta$ by a $\pi/2$ rotation), i.e. the acceleration is centripetal; its modulus reads:

$$\|\vec{a}\| = \omega R \dot{\theta}. \tag{14}$$

---

[11] $\vec{V}$ can be obtained analytically by calculating the time derivative of $\overrightarrow{SP} = R(1-e\cos\theta)\vec{u}_r$ with $d\vec{u}_r/dt = \dot{\theta}\vec{u}_\theta$ and $\dot{\theta} = \omega(1+2e\cos\theta)$ (from equations (3) and (5)). It leads to: $\vec{V} = \omega R\left(e\sin\theta\,\vec{u}_r + e\cos\theta\,\vec{u}_\theta + \vec{u}_\theta\right) = \omega R\left(\vec{e} + \vec{u}_\theta\right)$.

Using (6) to eliminate $\dot{\theta}$ and (4) to introduce $r$, it can also be written:

$$\|\vec{a}\| = \frac{\omega^2 R^3}{r^2} = \omega^2 R(1 + 2e\cos\theta). \tag{15}$$

If one supposes that the acceleration, in modulus, is a function $a(r)$ of the distance of the planet to the Sun, the dependence of $\|\vec{a}\|$ on the angular variable $\theta$ in equation (15) is entirely due to that of the distance $r = R(1 - e\cos\theta)$ when the planet explores its orbit. From the Taylor expansion

$$a(r = R(1 - e\cos\theta)) = a(R) - eR\cos\theta\, a'(R) + ..., \tag{16}$$

($a'(r) = da/dr$ being the derivative of $a(r)$), and its comparison with (15), one derives the two relations for the unknown function $a$:

$$a(R) = \omega^2 R \quad ; \quad \frac{a'(R)}{a(R)} = -\frac{2}{R}. \tag{17}$$

Since the equant model is supposed to be valid for any planet, therefore implicitly for any value of $R$, one deduces from the above differential equation that the acceleration $a(r)$ is proportional to the inverse square of the distance:

$$a(r) = \frac{\alpha}{r^2}. \tag{18}$$

In the above equation, $\alpha$ is a universal constant which a priori depends only on the attracting center $S$. Since also $a(R) = \omega^2 R$, the value of $\alpha$ is $\alpha = \omega^2 R^3$. This relation reads for the period $T = 2\pi/\omega$:

$$T = 2\pi\alpha^{-1/2} R^{3/2}. \tag{19}$$

Therefore the equant model, and the hypothesis that $a(r)$ is a universal function for the solar system, lead to Newton's $r^{-2}$ law for the acceleration and to Kepler's third law for the period, in the limit of small eccentricities.[12]

7. **Conclusion: Physics and the equant model.**

The above kinematical analysis of Kepler's equant model, shows that this model, which was already known by him to be a good approximation for the planet motions at the observational level, is the first order approximation (in eccentricity) of the Newtonian theory of gravitating bodies. In particular, it introduces two points (*S* and *Q*) which will be the focuses of the Keplerian orbit, it exhibits Kepler's second law of equal areas, Hamilton's hodograph and the centripetal acceleration; in addition, if one supposes that the acceleration is a universal function of the distance to the Sun, one recovers Newton's $r^{-2}$ law for the acceleration and Kepler's third law for the periods. This may allow us speaking of this model as a "pre-Newtonian" approximation (paraphrasing the post-Newtonian approximation of general relativity). As such, it is already interesting to introduce it as a kinematical exercise in the teaching of Mechanics. The equant model will then appear to students, no longer as an obsolete model or as an able foil to Kepler's final discoveries, but as a relevant contribution to the historical development of physics.

---

[12] One might deduce straightforwardly from equation (15) that if $a(r)$ is a universal function, this function is proportional to $r^{-2}$ and $\omega^2 R^3$ must be a constant. This is clearly a solution, but its unicity needs to be proved as we have done in the text.

There is another speculative but instructive way to look at the equant model and the previous calculations. If all the planets have had sufficiently small eccentricities so that Tycho Brahe's observations would have agreed with this model, it would have been the starting point for the future elaboration of physical laws, or at least for a reflexion on the nature of the influence of the Sun on the motion of the planets. If the Sun is responsible for their speed, the "velocity law" $V = Kr^{-1}$, with $K$ being a universal constant (for the solar system), is the natural candidate. This law is physically acceptable since it has a predictive power: the relation $K = \omega R^2$ implies that the period $T$ is proportional to $R^2$ (as seen in section 4). However, the experimental law is "$T$ proportional to $R^{3/2}$". This invalidates the hypothesis that $K$ is a universal constant. Certainly, one may "save the phenomena" by assuming, like Kepler, that the factor $K$ decreases with $R$, either because the driving efficacy of the Sun weakens with $R$, or because the resistances of planets to motion increase with $R$. But one may also abandon the "velocity law" and replace it by an "acceleration law". As we have seen in section 6, the generalization of the $r^{-2}$ law, valid for a planet on its orbit, to a universal one (Newton's law of gravitation), leads to the correct dependence of the period $T$ on $R$. Kepler was far from the formulation of such a hypothesis, since he was in an Aristotelian perspective of dynamics, where the velocity is proportional to the force. As well known, the acceleration has been recognized in physics as being a particularly relevant variable, only after Galileo's work on falling bodies (published eight years after Kepler's death), and Newton' association of it with gravitation. From our modern perspective, the equant model, which is already a good approximation for planetary motions, is also interesting because it allows a pedagogical discussion of the status of speed and acceleration in Mechanics.